\documentclass[useAMS,usenatbib]{mnras}
\usepackage{graphicx}
\usepackage{rotating}
\usepackage{aas_macros}
\usepackage{amsmath}
\usepackage{amssymb}
\usepackage{color}

\def\etal{{\it et al. }} 
\def\msun{{M$_\odot$}}

\title[An Upper Limit to Star Cluster Mass?] {Is There a Fundamental Upper Limit to the Mass of a Star Cluster?}

\author[Norris \etal] {Mark A. Norris$^{1,2}$\thanks{mnorris2@uclan.ac.uk}, Glenn van de Ven$^{2,3,4}$, Sheila J. Kannappan$^{5}$, 
Eva Schinnerer$^{2}$ $\&$ \newauthor Ryan Leaman$^{2}$
\\
  $^1$ Jeremiah Horrocks Institute, University of Central Lancashire, Preston, Lancashire, PR1 2HE, UK \\
  $^2$ Max Planck Institut f\"{u}r Astronomie, K\"{o}nigstuhl 17, D-69117, Heidelberg, Germany \\
  $^3$ European Southern Observatory, Karl-Schwarzschild-Str. 2, D-85748 Garching bei Munchen, Germany \\ 
  $^4$ Department of Astrophysics, University of Vienna, T\"urkenschanzstrasse 17, 1180 Vienna, Austria \\
  $^5$ Dept. of Physics and Astronomy UNC-Chapel Hill, CB 3255, Phillips Hall, Chapel Hill, NC 27599-3255, USA}
\begin{document}

\date{Accepted 2015 ***. Received 2015 ***; in original form ***}

\pagerange{\pageref{firstpage}--\pageref{lastpage}} \pubyear{2015}

\maketitle

\label{firstpage}

\begin{abstract}
The discovery around the turn of the millenium of a population of very massive (M$_\star$~$>$~2$\times$10$^6$ M$_\odot$) 
compact stellar systems (CSS) with physical properties (radius, velocity dispersion, stellar mass etc.) that are intermediate between 
those of the classical globular cluster (GC) population and galaxies led to questions about their exact nature. 
Recently a consensus has emerged that these objects, usually called ultra compact dwarfs (UCDs), are a mass-dependent mixture of 
high mass star clusters and remnant nuclei of tidally disrupted galaxies. The existence 
of genuine star clusters with stellar masses $>$10$^{7}$ M$_\odot$ naturally leads to questions about the upper mass limit of the star 
cluster formation process. In this work we compile a comprehensive catalog of compact stellar systems, and reinforce the evidence 
that the true ancient star cluster population has a maximum mass of M$_\star$$\sim$ 5$\times$10$^{7}$ M$_\odot$, corresponding
to a stellar mass at birth of close to 10$^{8}$ M$_\odot$. We then discuss several physical and statistical mechanisms potentially
responsible for creating this limiting mass.

\end{abstract}

\begin{keywords}
galaxies: star clusters, galaxies: formation, galaxies: evolution

\end{keywords}

\section{Introduction}

In the last wo decades the previously clear distinction between star clusters and galaxies
has been blurred by the discovery of new classes of stellar system. Particularly intriguing was
the unexpected discovery of a population of luminous, but compact, stellar systems which smoothly 
extend between the star cluster and galaxy sequences in various observational planes, such
as mass-size, effective surface mass density-mass, and velocity dispersion-mass \citep[see e.g.][]{Hasegan05,Kissler-Patig06,Misgeld11,Brodie11,AIMSSI}.
These objects, generally called ultra-compact dwarfs \citep[UCDs:][]{Minniti98,Hilker99,Drinkwater00,Phillips01} 
posed a major problem as they were not easily classifiable as either star clusters or galaxies. This led to much discussion over 
whether these objects were merely the high mass (and physically extended) tail of the normal globular cluster population 
\citep[e.g.][]{Fellhauer02,Mieske12}, or were in fact the remnant nuclei of dwarf galaxies tidally disrupted through 
interactions with larger companions \citep{Bekki03,Pfeffer13}. 

Additionally, over the same period a further complication arose that makes it even more difficult to 
separate bona-fide star clusters from galaxies and hence to determine which formation channel is 
responsible for creating UCDs. This was the discovery that most Milky Way GCs are not strictly 
true single stellar populations (SSPs), but in fact display complex abundance spreads \citep[see e.g.][]{Gratton12}. This discovery complicated
the use of one of the simplest discriminators between GCs and galaxies and raised questions about 
what the true definition of a star cluster or galaxy should be \citep{ForbesKroupa11,Willman12}.
For the purposes of this work we define galaxies as those objects located at the bottom of 
a potential well created by a combination of baryons and dark matter. This location means that they 
have the potential to acquire additional gas over time and can undergo repeated periods of star formation
and metallicity enrichment. Star clusters lack this privileged position and are therefore limited to forming 
stars using only the gas they are born from, or from any gas they can hold onto as it is released by stellar 
evolution. Therefore, their stellar populations are necessarily simpler, and their stars cannot, for example, 
display broad Fe-peak metallicity distributions seen in even the lowest mass Milky Way satellite galaxies 
\citep[see e.g.][]{Koch06,Starkenburg13,Hendricks14}.

Fortunately, based on significantly increased data samples, in recent years a consensus has begun to emerge 
that both suggested channels are responsible for forming UCDs \citep[][]{Hilker06,Norris&Kannappan11,Chiboucas11,Brodie11,AIMSSI,AIMSSII,Pfeffer14,Pfeffer16,Voggel16}.
This change was motivated by the observation that while the numbers of UCDs are in general in excellent
agreement with those expected from an extrapolation of the globular cluster luminosity function 
\citep[GCLF: ][]{Hilker06,Norris&Kannappan11,Mieske12}, an increasing number of cases of definitively stripped nuclei 
UCDs do exist \citep[][]{Norris&Kannappan11,Seth14,Norris15,Jennings15,Ahn17,Ahn18}. Furthermore, 
cosmological simulations indicate that stripped nuclei could make up a significant fraction of the UCD population only at the
highest masses ($>$10$^7$M$_\odot$), and should be a relatively negligible component ($<$ 10$\%$) at the lowest masses \citep{Pfeffer14}.

This realisation has led to a shifting of emphasis towards finding diagnostics to determine which route was at 
work for particular objects. It is relatively straightforward to classify some objects as former nuclei; if they are
still associated with stellar or gaseous debris streams \citep{Norris&Kannappan11,Jennings15,Schweizer18}, display
complex multicomponent structures or even their own associated GC systems \citep[][]{Hasegan05,Voggel16}, contain 
a supermassive black hole \citep{Seth14,Ahn17,Ahn18,Afanasiev18}, display extreme metallicities only found
in the central regions of galaxies \citep{AIMSSIII}, or exhibit an extended star formation history \citep{Norris15,Schweizer18}. 
For other objects no definitive signature of their origin might persist. For example, because the object is a 
hybrid; a true massive star cluster which through dynamical friction sank to the centre of a dwarf
galaxy to become its nucleus \citep[one of the proposed origins of such nuclei, see e.g.][]{Georgiev14}, 
and which was subsequently left behind when the surrounding galaxy was stripped by a tidal interaction \citep[see e.g.][]{Goodman18}.

As part of this effort, based on extrapolation of the empirically observed GCLF, \cite{Norris&Kannappan11} 
suggested the existence of an upper luminosity/mass limit for true star clusters. Given the properties of the GCLF; 
its approximately Gaussian shape \citep[see e.g.][]{Jordan07,Faifer11,Harris14}, universal turnover magnitude \citep[][]{Strader06}, 
and weak trend of increasing GCLF width with galaxy mass \citep[][]{Jordan07}, it is possible to estimate the 
luminosity of the brightest GCs expected to be found in a given GC system. This approach provides a remarkably
good match to the observed behaviour that the mass of the most massive GC in a GC system correlates strongly with
total GC system size \citep{Hilker09,Norris&Kannappan11}. \cite{Norris&Kannappan11} additionally found that 
given that the richest GC systems have around 10,000 - 20,000 members (those found around cD galaxies like M87), 
the most luminous GC-type UCD should have M$_{\rm V}$ $\sim$ --13, which for old stellar systems approximates 
to 7$\times$10$^7$ \msun. 

One important caveat to this argument is that the GC systems of galaxies are composite, 
built-up from GCs formed in-situ and those accreted from smaller companion galaxies \citep[see e.g.][]{Forbes10,Leaman13}. This implies 
that the total GC population available to produce a most massive GC will be less than that implied by
the present GC system size, as the accreted lower mass galaxies will not contribute particularly massive
GCs (due to the previously described trend of smaller GC systems having most massive GCs of lower mass). This effect has been observed
to have important implications, for example simulations indicate that it leads to the production of the ``blue tilt" observed 
in GC systems, whereby more massive GCs are on average redder and more metal rich \citep{Choksi18,Usher18}. This is explained by the fact that 
lower mass galaxies are lower metallicity and can only produce lower metallicity and lower mass GCs, meaning fewer low metallicity
GCs exist moving up the GC luminosity function, thereby changing the average GC metallicity as a function of GC mass.

Given the fact that many if not most GCs of massive galaxies are accreted, as evidenced by the fact that simulations indicate that
as much of 80$\%$ of the stellar mass of massive galaxies forms ex-situ and is later accreted \citep[][]{Oser10,Rodriguez-Gomez16,Clauwens18,Choksi19b},
we must reduce the effective GC system size, removing those GCs formed in
low mass galaxies, which cannot produce massive GCs. If the GC system formed in-situ around a cD galaxy is reduced to around 10,000 members, implying
a 50$\%$ accreted fraction (in line with typical accreted stellar mass fractions for massive galaxies: \citealt{Rodriguez-Gomez16,Qu17}), the predicted maximum GC luminosity drops to M$_{\rm V}$ $\sim$ --12.5, leading to a mass limit of
around 4 $\times$ 10$^7$ M$_\odot$. This value is consistent with the mass regime where \cite{AIMSSIII}
observe a transition in the metallicity distribution of GCs/UCDs, with objects more massive than a few $\times$ 10$^7$ M$_\odot$
exclusively displaying extremely high metallicities. It is also consistent with the 2 $\times$ 10$^7$ M$_\odot$ limit
above which \citet{Pfeffer16} find that Virgo and Fornax UCDs can be entirely explained by the expected number of stripped nuclei.
Hence we propose that there should exist a limiting mass for a genuine old GC of around 4 $\times$ 10$^7$ M$_\odot$.

With the advent of recent more comprehensive searches for UCDs it is now possible to revisit this prediction. This paper is organised
as follows; Section 2 describes the construction of a catalog of massive compact stellar systems, Section 3 examines the luminosity
function of CSSs for evidence of a truncation of true star clusters. Section 4 provides some suggestions for mechanisms which could be
responsible for creating the observed truncation, Section 5 provides a general discussion, and finally Section 6 provides some concluding
remarks.

\section{Catalog}

Until recently the principle problem limiting the study of massive compact stellar systems was a historic preference in
studies of GC systems to enforce either an upper magnitude or a size limit on the selected GC candidates, in order 
to reduce contamination from background galaxies. The relaxation of these limits (in order to allow UCDs into the selection), 
along with very deep spectroscopic surveys which are typically complete down to M$_{\rm V}$ $\sim$ --11  
\citep[e.g.][]{Mieske04,Misgeld11,DaRocha11,Mieske12}, and systematic searches for exactly the type of objects 
previously excluded \citep[see e.g.][]{AIMSSI} has allowed for the compilation of large catalogs of compact stellar 
systems spanning the GC to galaxy regimes. 

While these catalogs are by no means homogeneous, or complete, especially at low luminosities/masses, the fact that 
the most extended and luminous objects are the easiest to find and spectroscopically confirm ensures that the census 
of massive UCDs is close to complete for the area surveyed. Therefore, until truly volume-limited spectroscopically-confirmed 
samples selected from surveys such as the Next Generation Virgo Cluster Survey (NGVS: \citealt{NGVS1}) become available, 
these compilations remain the most comprehensive.

In this work we compile the most extensive catalog of spectroscopically confirmed CSSs, in order to search for a 
truncation in the upper mass of star clusters. The principle sources for the catalog are the previous compilations of 
\citet[][]{Brodie11}, \citet{Misgeld&Hilker11}, and in particular the Archive of Intermediate Mass Stellar Systems \citep[AIMSS:][]{AIMSSI,AIMSSII,AIMSSIII}.
These compilations include the Coma Cluster UCD's of \citealt[][]{Chiboucas11}, the Perseus Cluster UCD sample of \citealt{Penny12,Penny14}, 
the Antlia Cluster UCD sample of \citealt[][]{Caso13,Caso14}, the Centaurus A UCDs of \citealt[][]{Taylor10}, and the NGC~1132 UCD's of \citealt[][]{Madrid13}. 

To these catalogs we add additional M87 UCDs from \cite{Hasegan05,Zhang15}, the UCD of NGC~5044 \citep{Faifer17}, the UCD of NGC~7727 \citep{Schweizer18},
GCs from the Milky Way system \citep[][2010 edition]{Harris96}, M31 GCs \citep{BolognaM31GCs}, and GCs of the 
Hydra I cluster \cite{Misgeld11}. Finally we include the sample of GCs detected in the ACS Virgo Cluster Survey 
\citep[ACSVCS: ][]{ACSVCSI,Jordan07}, this sample is not spectroscopically confirmed, but due to the excellent HST 
imaging, contamination of the high-confidence GC sample (we select only objects with GC probability $>$ 95$\%$) is expected to be negligible. 

Ideally we would examine the mass of the CSSs directly. However, due to the extreme inhomogeneity of the
available photometry this is not possible. We therefore examine the distribution of absolute V magnitudes, as these
are most readily available in the literature and are a good proxy for stellar mass for old stellar systems. The only
limitation we impose is to exclude the handful of CSSs with spectroscopically derived ages $<$ 3 Gyr, so that younger CSSs
 do not appear artificially bright when compared to the majority older 
population. This removes only a handful of young clusters from nearby merger systems 
(such as NGC~7252), plus a few suspected stripped-nucleus type UCDs.

\section{Results}

\begin{figure} 
   \centering
   \begin{turn}{0}
   \includegraphics[scale=0.52]{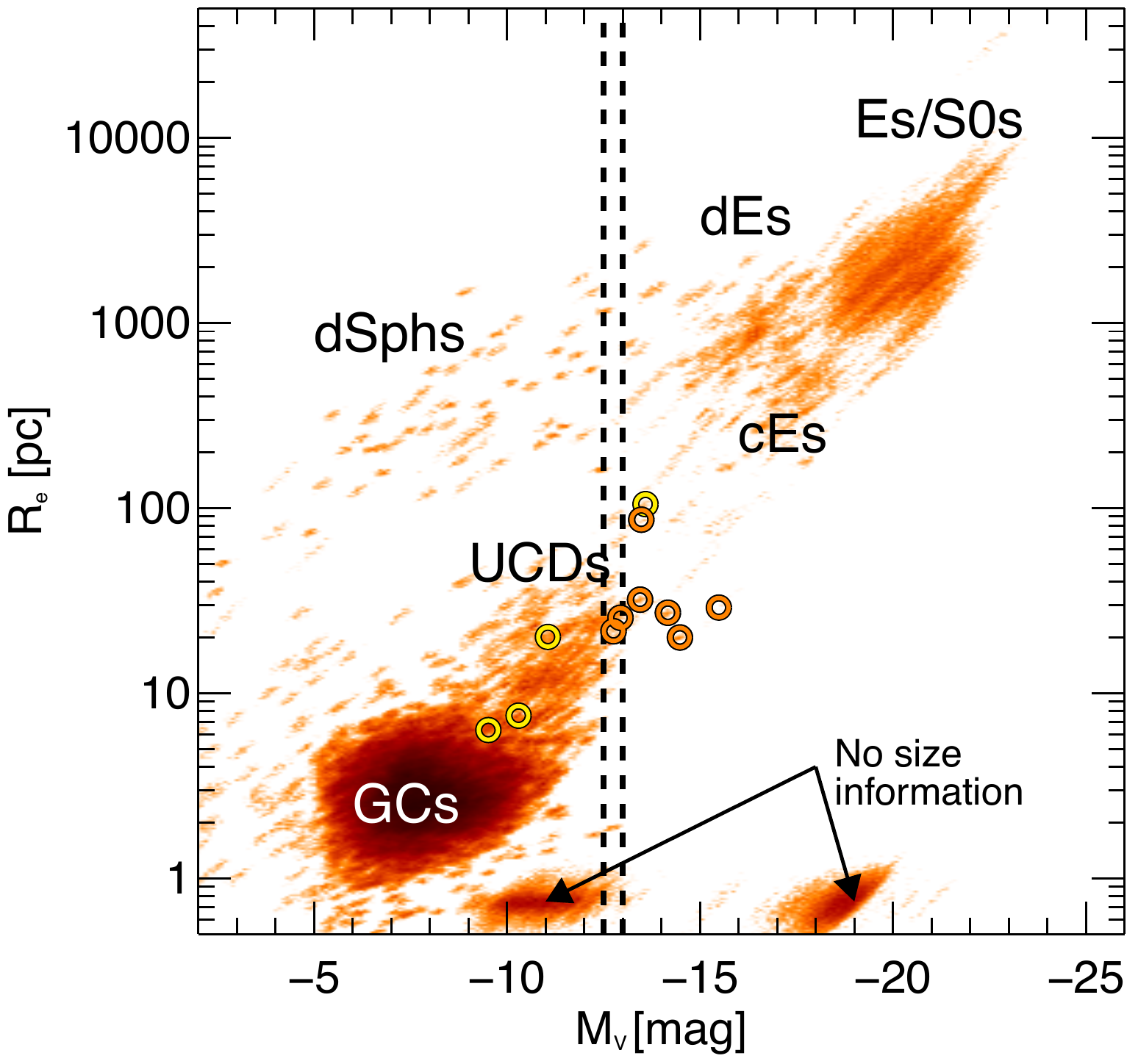}
   \end{turn} 
   \caption{The luminosity-size plane for dynamically hot stellar systems. Rather than plotting points
   we show the probability density for each object, by including the uncertainties on the distance, size
   and magnitude. This more accurately reflects the inherent correlations in the absolute magnitude
   and radius (caused by their mutual dependence on distance). The clouds of objects with R$_{\rm e}$ $\sim$0.75 pc 
   and M$_{\rm V}$ $\sim$ --11 and --19 are those UCDs and cEs which have no measured size and are 
   therefore given arbitrary size. The six orange open circles indicate those
   UCDs known to be stripped nuclei (M60-UCD1, NGC~4546-UCD1, M59cO, VUCD3, UCD3, M59-UCD3, 
   NGC~7727-Nucleus~2), the yellow open circles are highly suspected stripped nuclei ($\omega$ Cen, M54, 
   S999, VUCD7). The vertical dashed lines show the proposed region between M$_{\rm V}$ = --12.5 and --13 mag 
   where star clusters cease to exist. The dramatic drop-off in numbers of objects in this luminosity range is clear.      }
   \label{fig:lum_size}
\end{figure}

Figure \ref{fig:lum_size} shows the location of our CSS sample, plotted in the luminosity-size plane. Other dynamically hot 
stellar systems are also plotted for illustrative purposes. 
This plot shows that despite the fact that objects tend to scatter diagonally (due to common dependence of the absolute magnitude 
and physical effective radius on the distance estimation),
approximately along the line connecting GCs and galaxies, very few objects are consistent with being more luminous than
M$_{\rm V}$ $<$ --13 and more compact than R$_{\rm e}$ $\sim$ 200 pc. Furthermore, the 7 UCDs which are
unambiguous stripped nuclei (M60-UCD1, NGC~4546-UCD1, M59cO, VUCD3, UCD3, M59-UCD3, NGC~7727-Nucleus 2 indicated by orange circles)
are all broadly consistent with being M$_{\rm V}$ = --13 or brighter. The remaining objects more luminous than 
M$_{\rm V}$ = --13 either have not yet been studied in detail, or have no definitive evidence to prove their type either way.

We omit an examination of the cE population, which the UCD population may overlap with somewhat, due to their ambiguous origin, 
and the fact that they are unambiguously galaxies not massive star clusters. While it seems clear that many cEs are the result of tidal 
stripping interactions \citep{Huxor11b}, there is also the possibility that there may also be a population of intrinsically compact elliptical 
galaxies, analogous or related to the massive compact galaxies observed at higher redshift \citep[see e.g.][]{Kormendy09,vanderWel14}.
Even within the stripping scenario a diverse range of objects may result depending on whether the stripped galaxy is gas-rich or already
quenched. Going forward, our references to stripping formation scenarios should be interpreted to include the gas-rich dwarf accretion 
scenario of \cite{Du18}, where the dense metal-rich cE (or potentially UCD) is formed during the stripping event, by the ram pressure
confinement of the gas (and resulting rapid enrichment) of a central starburst triggered by the interaction.

 \begin{figure} 
   \centering
   \begin{turn}{0}
   \includegraphics[scale=0.52]{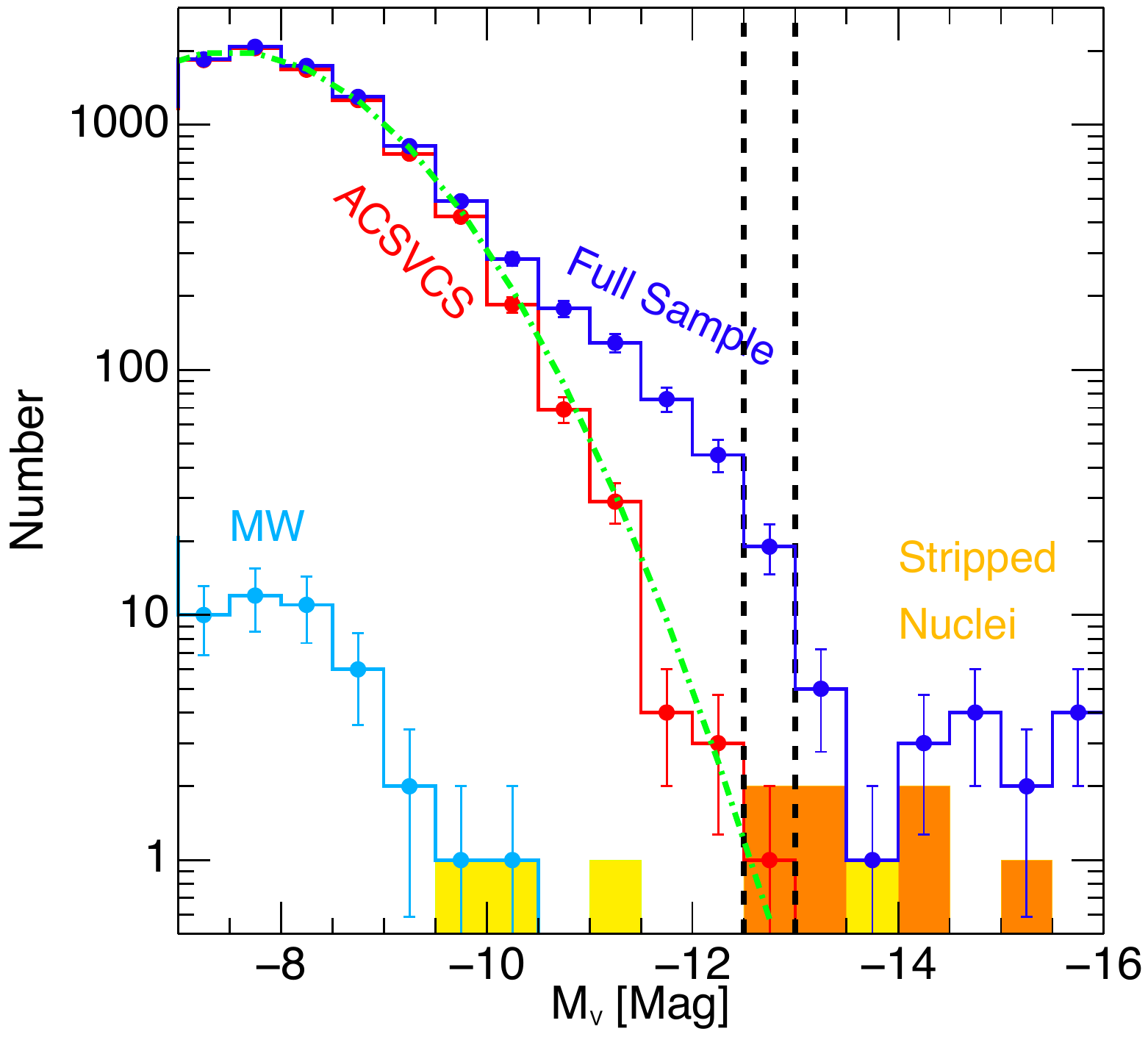}
   \end{turn} 
   \caption{Histogram of M$_{\rm V}$ for samples of CSSs. The blue histogram
   shows the full catalog of CSSs. The red histogram shows all objects from
   the ACSVCS which have $>$ 95$\%$ probability of being GCs or UCDs. The cyan 
   histogram shows the distribution of magnitudes of Milky Way GCs. The solid orange regions
   show the locations of those CSSs known to be stripped galaxy nuclei, the solid yellow regions indicate
   suspected stripped galaxy nuclei. The green dot-dashed
   curve is not a fit to ACSVCS, but shows a Gaussian with central M$_{\rm V}$ 
   = --7.5 mag, dispersion = 1.3 mag, and peak of 2000 GCs. The vertical dashed lines again show the 
   proposed region between M$_{\rm V}$ = --12.5 and --13 mag where true star clusters cease to exist, 
   above M$_{\rm V}$ = --13 mag the number of CSS's is consistent with being constant.
   }
   \label{fig:lum_hist}
\end{figure}

Figure \ref{fig:lum_hist} shows the magnitude distributions of various subsamples of CSSs. The blue histogram is the
full catalog compiled here, and although the constituent surveys all have differing selection criteria, they are all
fairly complete for objects brighter than around M$_{\rm V}$ = --11 or -11.5. The red histogram is the 
distribution of GC luminosities found in imaging of 100 Virgo cluster galaxies by the ACSVCS. It can be confidently 
assumed that if the regions around each of the non-ACSVCS CSSs were surveyed to the same depth as the ACSVCS, 
the blue histogram would assume an almost identical shape to that of the ACSVCS sample. The green 
dot-dashed line is \textit{not} a fit to the red ACSVCS histogram, but instead shows a Gaussian with mean magnitude and 
dispersion $\sigma$ chosen to match those found for the GC population of M87 (M$_{\rm V}$ = --7.5 and 1.3 respectively) and then arbitrarily 
normalised to match the ACVCS distribution (see \citealt{Liu15} for a similar examination of the GC/UCD population of M87). The solid orange histogram shows the magnitudes of 7 confirmed stripped 
nuclei-type CSSs, the solid yellow histogram shows the 4 strongly suspected former nuclei, and the dashed vertical black 
lines delineate the regime where true star clusters are proposed to cease to exist.

The close agreement between the model Gaussian and the ACSVCS histogram demonstrates that the CSS distribution 
for Virgo is well fit by a single luminosity function where the fit parameters are dominated by GCs within a few magnitudes 
of the turnover magnitude (as is typically the case for estimations of the GCLF of galaxies). The agreement between 
the upper limit where the green line predicts only a single star cluster, and our suggested upper magnitude limit for star cluster 
formation is by construction. As discussed in the introduction it was the observation that even GC systems with $>$10,000 - 20,000 
members would not predict more than $\sim$1 GC with magnitude $\lesssim$ --12.5 - 13 that motivated the definition of the upper limit. 

The full sample further supports our previous suggestion that M$_{\rm V}$ $\sim$ --13 mag marks a transition in the CSS population. 
There are 19 CSSs in the magnitude bin --12.5 $<$ M$_{\rm V}$ $<$ --13, but above this the number of 
objects is approximately constant with only $\sim$ 3 per 0.5 mag bin. This levelling-off in the number of objects 
more luminous than M$_{\rm V}$ = --13, despite them being more easily discovered, is evidence for a change in CSS behaviour 
at this magnitude. To definitely demonstrate the exact value of the transition magnitude will likely require the assembly of true
volume-limited and highly complete samples of CSS's, such as those assembled by combining deep imaging surveys such as the 
NGVS \citep{NGVS1} with equally complete spectroscopic follow-up.

Our interpretation of this behaviour, that above M$_{\rm V}$ = --13 all objects 
are stripped galaxy substructures, is further supported by the observation that seven of the objects with M$_{\rm V}$ $\lesssim$ --13 
have already been shown to be ex-nuclei \citep[e.g.][]{Seth14, Norris15, Ahn17}. We predict that the majority of objects more luminous
than M$_{\rm V}$ $<$ --13 will display unambiguous evidence of a galactic origin (some formed by stripping at early epochs may 
be indistinguishable from star clusters in practice). 

From our examination of the stellar masses of CSSs \cite[see e.g.][]{Norris&Kannappan11,AIMSSI,AIMSSIII} we found that the 
suggested M$_{\rm V}$ = --13 mag limit translates into a current stellar mass of around 3-7$\times$10$^{7}$ 
\msun\, for these objects. Assuming a Kroupa initial mass function \citep{KroupaIMF} the stellar mass loss due to stellar evolution over 10 Gyr
is around 30$\%$ \citep{Into13}. Therefore at birth our limit translates to a maximum stellar mass for a true stellar cluster
of between 7$\times$10$^{7}$ and 10$^8$ \msun, depending on the fraction of gas from stellar evolution that the cluster can retain.

A final piece of evidence in favour of the proposed scenario is provided by the observation of young massive star clusters in nearby 
galaxies. To date, the most massive young star cluster discovered is NGC~7252-W3, which has a mass of (8 $\pm$ 2) $\times$ 10$^{7}$ \msun\, and
a radius of 17 pc at an age of around 500 Myr \citep{Maraston04}. This cluster, along with all other young massive clusters which 
approach the proposed limit are found associated with ongoing or recent major galaxy mergers, perhaps indicating that unusually violent
events are required to create such massive star clusters \citep[see also][]{Bastian13}, at least at z=0.

\section{Causes of the Upper Mass Limit}
 
Having demonstrated that the luminosity function of CSSs supports the existence of an upper mass limit for true 
star clusters of around 10$^8$ \msun, we now consider the mechanisms potentially 
responsible for creating the limit.

\subsection{Scenario A: The Need for Extreme ISM Densities and Pressures}
\label{Sec:ISM}

\cite{Kruijssen12} presents a theoretical scenario in which gravitationally bound star clusters form across the
density spectrum of the ISM, but with increasing efficiency at higher densities. This leads naturally to the
prediction that to form the most extreme members of the star cluster population requires extreme conditions 
in the ISM. Unfortunately there have been relatively few simulations that can directly test such a scenario, as
the required simulations must capture massive star cluster formation across the full range of ISM conditions found
in galaxies or galaxy mergers. In part this lack of suitable simulations is because until recently it has been technically 
impossible to adequately sample the range of spatial scales involved, as galaxies and mergers typically require
examination across tens of kpc, while GCs have half light radii of only $\sim$2 pc.

However, \cite{Renaud14,Renaud15} extending the work of \cite{Bournaud08} presented a hydrodynamical 
simulation which attempts to reproduce the well studied ongoing merger system of the Antennae galaxies. This
simulation has resolution of 1.5 pc and includes star formation and stellar feedback, allowing a more detailed examination
of the properties of young clusters formed in the merger. By comparison with a similar simulation of an isolated Milky
Way-like galaxy from \cite{Renaud13} they are able to contrast the properties of compact stellar systems formed in 
relatively quiescent galaxies, versus those formed in intense merger induced starbursts.

The headline result from the \cite{Renaud13,Renaud14,Renaud15} simulations are that the MW simulation does
not form any star clusters more massive than 3 $\times$ 10$^6$ \msun, while the Antennae simulation 
creates star clusters up to a maximum mass of around $\sim$10$^8$ \msun\, with radii of 10 to 30 pc, similar to
those of UCDs or extreme young massive clusters such as NGC~7252-W3. They find that 
star clusters up to 5$\times$10$^7$ \msun\, form in or close to the tidal tails, and clusters of up to 10$^8$ \msun\,
form in the densest central regions during the final coalescence \citep[see also][for simulations that create massive 
central star clusters of mass $\sim$10$^8$ \msun]{Li04,Matsui12}. They conclude that the galactic interaction leads 
to tidally and turbulently compressive regions in the ISM which in turn leads to the formation of clusters 30 
times more massive than those found in quiescent discs. One caveat to this work, is that the most
massive star clusters formed in the simulations often display an age spread of up to 100Myr, due to ongoing 
accretion of gas leading to prolonged star formation. Such extended star formation histories are ruled out for
for modern young massive clusters of mass 10$^6$ - 10$^7$\msun\, \citep[see e.g][]{Cabrera-Ziri14,Cabrera-Ziri15},
but it is currently not possible to place stringent limits on the length of star formation for any of the bona-fide
massive UCDs of our sample, due to their distance, and hence unresolved stellar populations. 

Nevertheless, more recent studies find broadly similar results to the studies of \cite{Renaud13,Renaud14,Renaud15}. 
For example, when studying the formation
of bound stellar clusters in simulated interacting galaxies \cite[][]{Maji17}  find that clusters as massive as 10$^{7.5}$ \msun\,
can be formed, but they form preferentially in the most highly-shocked regions of galaxy interactions where the 
pressure is 10$^4$ -10$^8$ times larger than typical for the ISM. Similarly, based on high resolution simulations 
\cite{Ma19} find that bound clusters form preferentially in high-pressure, high-density environments, and further
suggest that external pressure (from colliding clouds/gas streams or feedback winds) is required to produce the necessary
pressures to form proto-GCs. Other studies have likewise found that merger 
induced interactions may be required to produce sufficiently high pressures and densities to produce star clusters 
significantly above the typical turnover mass for GCs of 2 $\times$ 10$^5$\msun\, \citep[see e.g.][]{Li17,Kim18}. 

In conclusion it seems that in order to form massive star clusters it is necessary to have very high gas densities 
with significantly higher compression (due to turbulence) than is present in present day quiescent discs. However, 
it is clear that the necessary gas densities and turbulence appear to 
have been much more common at higher redshift when the bulk of the massive star cluster population was 
formed, even in the so-called clumpy discs commonly observed at higher-z \citep[see e.g.][]{Swinbank11, Falgarone17}.
It is also interesting to note that significant samples of objects with masses (10$^6$ - 2 $\times$ 10$^7$ \msun) 
and sizes expected of the progenitors of modern massive star clusters are beginning to be resolved in studies of 
lensed galaxies at z = 3 - 8, exactly when the bulk of GC progenitors are expected to 
form \citep{Vanzella17,Bouwens17}.

The observation that very high gas densities and pressures are required to form massive star clusters naturally 
leads to a limitation on the maximum mass of a cluster that can form (see also
\citealt{Elmegreen18} for similar arguments relating to the formation of GCs in high redshift galaxies). Even in such a large merger as the Antennae the physical conditions 
never reach the threshold required to form clusters of 10$^8$ \msun,
except in the very central regions of the merger, where any clusters that form are quickly incorporated into
the bulge. For star clusters formed in the tidal tails (i.e. those formed on orbits that could allow them to survive 
for a Hubble time) the maximum compression is necessarily lower than that reached in the central regions at 
the bottom of the galactic potential. This is because the gas in tidal tails can expand outwards perpendicular 
to the gas inflow along the tidal tail, whereas in the central regions additional infalling gas can keep the gas 
pressure high allowing higher mass clusters to form.

\subsection{Scenario B: Insufficient Molecular Gas}
\label{Sec:MolecularGas}

One obvious observation regarding the existence of an upper mass limit for star cluster formation is that 
the total stellar mass formed should be significantly higher than that of the most massive cluster. This is
because young star clusters of the type likely to evolve into GCs and UCDs do not generally form alone, 
but in fact form in large numbers during violent galaxy interactions \citep[see e.g. the YMC populations
of NGC~1316 and NGC~7252][]{Goudfrooij12,Bastian13}. From Section \ref{Sec:ISM} we see that we
expect that the most massive star clusters form in major galaxy interactions where the ISM density
and turbulence is high. We also expect major galaxy mergers to be a site of formation for massive 
clusters because sufficient quantities of gas are available, and the star clusters can form on orbits that keep them 
away from the galaxy centre or disc, which protects them from total disruption through dynamical friction 
within a short period, allowing them to survive until the present epoch.

Observations of young star clusters in merger systems show that the number of clusters follows a power 
law dependence on cluster mass of the form $dN / dM \propto M^{-\beta}$ with $\beta$ = 2 \citep{Fall09}. 
Furthermore, observations indicate that the most massive star clusters found in these 
nearby merger remnants are consistent with the expectations of simply statistically sampling from the 
same power law as the bulk cluster population \citep[see e.g.][for the cases of NGC~34, NGC~1316, NGC~3610, NGC~4038/39 and NGC~7252 respectively]{Schweizer07,Goudfrooij04,Whitmore02,Whitmore10,Miller97}.

As \cite{Elmegreen12} demonstrate, this mass dependence can be reformulated to provide a prediction
for the total stellar mass formed in a given star forming period that produces a cluster of mass $M$ 
(their equation 2):

\begin{equation}
\begin{split}
M_{\rm total} = \eta^{-1}_c M + \eta^{-1}_c(\beta -1) M^{\beta - 1} \\
\times (\frac{M^{2 - \beta } - M^{2 - \beta}_{\rm min}}{2 - \beta}, ln[\frac{M}{M_{\rm min}}])
\end{split}
\end{equation}

\noindent Here $\eta_c$ is the fraction of stars formed in star clusters. Following \cite{Elmegreen12} we
assume a conservative fraction of 0.25, as the value of $\eta_c$ has been claimed to vary significantly with
local physical conditions \citep{Silva-Villa13}. \citealt{Adamo15b} suggest that the fraction of stars forming in
bound clusters varies from $\sim$3$\%$ in quiescent dwarf galaxies, to $\sim$50$\%$ or more in
the most intense starbusts. \citealt{Kruijssen12} likewise suggest a range of between $\sim$3$\%$ in the 
lowest density galaxies and 70$\%$ in the highest density systems. Finally, the E-MOSAICS simulations \citep{EMOSAICS}
indicate cluster formation efficiencies that are generally 20-30$\%$ at z=6, can reach $\sim$80$\%$ during bursts
at intermediate redshifts, and decline to $\sim$ 1$\%$ at z=0 for their simulated Milky Way analogues.

We note however, that recently \cite{Chandar17} claim that the apparent variation in the fraction of stars forming in
clusters with environmental conditions is in fact due to observational inconsistencies, and
in particular the fact that the times since cluster formation is different in each case, leading to differing
amounts of cluster dissolution. 

Nevertheless the value for the fraction of stars formed in clusters that they determine
(24 $\pm$ 9$\%$) is consistent with our adopted value. M$_{\rm min}$ is the minimum bound cluster mass, 
which in line with \citealt{Elmegreen12} is assumed to be 10 \msun\,here. 
 
Note that in using this derivation we explicitly assume that there is no physical truncation of the initial cluster
mass function, clusters can form up to any mass, as long as sufficient gas is available. This is in contrast to
a cluster mass function of the form typically found for disc galaxies, where an exponential truncation \citep{Schechter76}, 
generally occurs at a few $\times$ 10$^{5}$ \msun\, \citep{Gieles06,Bastian08,Gieles09,Larsen09,Kruijssen14,Adamo15,Adamo17}. 

Using this equation we find that in order to form a most massive cluster with mass M = 10$^{8}$ \msun\, 
the total mass of stars formed in the star formation event is 7$\times$10$^9$ \msun. We note 
that these values are consistent with those found by the E-MOSAICS simulations \citep[see figure 5 of][]{EMOSAICS}.
They are also broadly consistent with the findings of \cite{Ma19}, who find that to form a cluster of mass M$_{cl}$ requires
the formation of 20 M$_{cl}$ of stars in the galaxy as a whole. 
As the total efficiency of the conversion of molecular gas to stars ($\eta_{\star}$) is never unity, the total amount of 
molecular gas required will be significantly larger. Assuming the average star formation efficiency is similar
to that observed in Milky Way molecular clouds  \citep[i.e. 2$\%$;][]{Leisawitz89}, the total molecular 
gas required would be of the order 4$\times$10$^{11}$ \msun. 
Alternatively, the required molecular gas mass could be reduced by a factor of 10-20 if the
star formation efficiencies were assumed to be in the range thought to be required for a star 
cluster to remain bound after gas expulsion \citep[i.e. SFE $>$ 20-40$\%$][]{Parmentier08,Smith11}. 
Such high average star formation efficiencies are observationally motivated, as observations of starbursts 
indicate that they are forming stars more efficiently than local spiral discs by factors of $>$ 10 \citep[see e.g.][]{Meier10,Silverman15}

Therefore, we arrive at a required molecular gas mass in the range of $\sim$1$\times$10$^{10}$ to 
4$\times$10$^{11}$ \msun\, in order to form a most massive star cluster with stellar mass of 
10$^{8}$ \msun\, at birth. Note that this molecular gas mass does not necessarily all have to be concentrated 
in a single star forming complex, but at least 10$^{8}$ $\times$ $\eta_{\star}^{-1}$ \msun\, must be located 
within a single bound structure to create the most massive cluster.

The need for such enormous quantities of cold molecular gas naturally places a strong constraint
on the probability of forming such massive clusters. Even massive disc galaxies, such as the Milky Way or M51, 
typically have total molecular gas masses of only around 5$\times$10$^{9}$ \citep{Shetty07,Schinnerer13}
and no single cloud has a mass that exceeds 2$\times$10$^{7}$ \msun\, \citep{Colombo14}. In fact, 
in a study cross-matching ALFALFA and SDSS data of $>$ 11,000 galaxies out to $z$ = 0.06 \cite{Maddox15} find
few galaxies with cold gas mass $>$ 10$^{10}$ \msun\, and none with $>$ 10$^{11}$ \msun.
In contrast, at $z$ $>$ 1.5 cold gas masses of $>$ 10$^{11}$ \msun\, are seen, at least for the most massive 
galaxies thought to be likely progenitors of early-type galaxies \citep[see e.g.][]{Tacconi12,Scoville16,Rudnick17}.
However, no galaxies with gas mass $>$ 10$^{12}$ \msun\, are seen at $z>2$ (or anywhere else), 
despite being easier to detect. It is therefore plausible that the lack of star clusters with mass $>$ 10$^{8}$ \msun\, could be down to 
the fact that there simply aren't any galaxies/mergers where sufficient cold gas is available at any one 
time to create them.

This scenario is therefore statistical in nature; there needn't be a physical limitation of gas physics which 
prevents larger clusters forming, it is simply that our Universe rarely, if ever, brings together enough cold gas to
create such clusters, leading to a practical limit on the maximum cluster mass found in a reasonable volume. It might therefore 
be speculated that the upper limit produced by this scenario is the result of cosmology, with the interplay of the initial matter 
power spectrum and the expansion rate of the Universe ultimately setting how much gas can be accumulated at any 
one epoch.

\subsection{Scenario C: Shear}

\cite{Reina-Campos17} present a simple analytical model to determine the
maximum mass of star clusters. They suggest that the limiting mass is set by a combination
of stellar feedback and environmental shear. Their results indicate that cluster formation
within Milky Way-like spiral discs will typically be feedback limited beyond 4 kpc, while more massive higher 
redshift star formation will likely be shear limited at all radii, a change driven by the large increase in 
gas surface density at higher redshift. Their model predicts maximum star cluster masses
that broadly agree with our limit at high redshift ($\sim$10$^8$-10$^9$ \msun), with a reduction 
to $\sim$ 10$^4$ - 10$^5$ \msun\, for local galaxy discs. However, this model assumes cluster 
formation occurs within a differentially rotating disc in hydrostatic equilibrium, an assumption
that has been shown to be valid for high-redshift galaxies because despite the clumpy and 
chaotic nature of young galaxies, simulations indicate that star formation is still restricted to 
relatively thin disks \citep[][]{Meng19}. However, it remains unclear how
reliable an assumption this would be for the progenitors of today's GCs and UCDs, as these objects
are now found on orbits that keep them well away from the densest regions of their host galaxies,
where dynamical friction would rapidly lead to their destruction. Presumably an interaction would be 
required to eject the proto-star clusters onto orbits with longer dynamical friction timescales,
 it is not currently clear whether such interactions would be common enough to explain the observed 
 abundance of massive star clusters.
Nevertheless, this model if extended to describe the shear environment of a major merger 
or starburst holds promise for explaining the maximum star cluster mass across all mass scales.

\subsection{Scenario D: Stellar Feedback}

Stellar feedback alone may be able to explain the maximum star cluster mass, subject to 
some uncertainty regarding star formation efficiencies. Massive young stars emit copious amounts of high energy photons that 
deposit momentum into the surrounding ISM, when this exceeds the force of gravity the
gas is expelled and any further star formation is curtailed. As discussed in, e.g., \cite{Murray10,Hopkins10,Rahner17,Crocker18,Grudic19}
this behaviour is analogous to the Eddington limit for stars. However, one important difference
is that the ISM of star forming regions is dusty, and the opacity of dusty gas is much higher
than the electron scattering opacity found in stars. Therefore radiation from young massive
star clusters could efficiently act to restrict their own growth. 

When examining a range of dense stellar systems from GCs to galaxy spheroids, \cite{Hopkins10} 
find a nearly constant maximum central stellar surface mass density. They attribute this
maximum surface mass density to stellar feedback reaching an Eddington-like limit that 
regulates the growth of dense star forming regions. They also show that for certain assumptions
this Eddington-like limit is reached for a gas surface mass density of $\Sigma_{\rm gas}$ =
10$^{11}$ - 10$^{12}$ \msun\,kpc$^{-2}$. Recently \cite{Crocker18} confirm this result as 
being consistent with that expected to be caused by direct and indirect radiation pressure from 
the young stellar cluster. Converting this apparent limit into the correct area for 
typical massive UCDs (which have R$_{e}$ $\sim$ 20 - 100pc) and assuming the efficiency of 
gas to stellar mass conversion described in Section \ref{Sec:MolecularGas} (i.e. 2 - 40\%) does 
in fact produce stellar masses in the correct range for the most massive UCDs ($>$10$^7$ \msun). 

More detailed simulation work is required to constrain the expected range of star cluster 
formation efficiencies for this scenario (efficiencies as high as 90\% are found in the radiation limited
case by \citealt{Crocker18}), and to include other sources of energy injection (e.g. prompt SN) to see how
these will impact the final bound cluster mass. It may also prove the case that even when this
effect operates the maximum cluster mass is still limited by one of the other scenarios outlined,
for example by limitations on the availability of sufficient gas. {Recently \citealt[][]{Grudic19} 
have taken steps in exactly this direction to produce a modified model in which the upper limit of the 
stellar surface density is caused by stellar feedback becoming \textit{ineffective} above some critical 
threshold, thereby causing the supply of gas to be rapidly expended before the system can contract to
higher density.

\section{Discussion}

All four proposed mechanisms currently provide plausible explanations for why star clusters would experience
a maximum mass limit. Additional observational and simulation work will 
be required to determine which (if any) is responsible for the observed upper mass limit.

The first two mechanisms are essentially statistical in nature. This is a strength as it means
that they can potentially naturally explain not just the existence of the upper mass limit, but also the distribution 
of masses of star clusters. For scenario A a distribution of pressures and densities throughout the
merger leads to a range of initial star cluster sizes and masses. For scenario B a lower total galactic 
gas mass populates less far up the star cluster mass function, but still forms clusters up to that mass, 
and furthermore could produce the right mass function for star clusters, assuming the correct
GMC mass function and differential survival of YMCs to become GCs based on their mass.

This statistical nature is also a problem as in order to determine their efficacy in
producing the observed limit the observational or simulated data must be more comprehensive. 
For example, to accurately test mechanism A (extreme ISM conditions) will require simulating many different galaxy mergers 
and starbursts with the resolution (ideally more) used by \cite{Renaud14,Renaud15}, followed by 
comparing the produced mass distributions of surviving star clusters with those observed in massive early-type 
galaxies. Likewise, for mechanism B (insufficient gas supply) it is necessary to examine the molecular gas reservoirs and star formation
activity of a large ensemble of simulated galaxies, and to examine how the mass function of surviving
star clusters correlates with the available gas reservoirs at the epoch when they formed. 

Mechanism C (shear) is potentially similar, in that the correct distribution of stellar feedback and shear 
could lead to both the observed mass function of star clusters, and their ultimate upper mass limit. 
Simulations similar to those required to investigate mechanism A, plus observations of interacting 
and quiescent galaxies will eventually demonstrate whether the range of stellar feedback and 
shear environments present in such galaxies matches those required to explain the full mass range 
of star clusters. In the near future the ongoing simulations of the E-MOSAICS project \citep{Pfeffer18} which 
incorporate the formation and evolution of star cluster populations following the prescriptions of \cite{Kruijssen12} and 
\cite{Reina-Campos17} into the EAGLE simulations of galaxy formation \citep{Schaye15,Crain15} should 
demonstrate the efficacy of this scenario.

Further work is necessary to demonstrate if mechanism D (stellar feedback) can produce not only a maximum upper
mass limit, but also explain the observed mass function of star clusters, either alone or in combination with
one of the other scenarios.

Finally it is worth noting that the mechanisms described here apply only to the genuine star cluster population 
of UCDs. Those UCDs formed by the liberation of galaxy nuclei during tidal interactions would not be expected to be limited 
by any of the processes outlined, principally because galaxy nuclei can undergo repeated bursts of star formation 
\citep[see e.g.][]{Norris15}. This ensures the expected mass function of former-nuclei should extend to significantly higher
mass than that of the genuine star cluster population, and the presence of any upper mass limit to such objects becomes
difficult to discern due to overlap with other similar objects like compact ellipticals.

\section{Conclusions}

We have assembled the most comprehensive sample of compact stellar systems yet currently robustly classified. This
sample is the largest available compilation in the intermediate luminosity/mass regime, where the division between 
star clusters and galaxies is most uncertain. 

Using this catalog we have strengthened the existing evidence for the existence of an upper initial mass 
limit of surviving genuine star clusters at birth of around 10$^{8}$ \msun. Definitively demonstrating the exact
location of this limit will require future volume-limited, spectroscopically-confirmed and highly-complete 
CSS surveys.

We have examined four possible mechanisms responsible for the lack of bona-fide star clusters
with stellar masses $>$ 10$^8$ \msun, and conclude that all are plausible. Further simulation work 
looking at the ensemble properties of the galaxy population at higher redshift (to examine
the cold gas distributions), additional higher resolution simulations of major mergers (to check the distributions
of most massive clusters produced), and focussed simulations
of single massive star cluster formation (to examine the effect of stellar feedback) will be required to
determine which is the principal effect.

\section{Acknowledgements}

The authors would like to thank the anonymous referee along with Sharon Meidt and Brent Groves
for their extremely helpful comments that greatly improved this paper.
 
 GvdV acknowledges funding from the European Research Council (ERC) under the European Union's Horizon 2020 research and innovation programme under grant agreement No 724857 (Consolidator Grant ArcheoDyn).

\bibliographystyle{mnras}
\bibliography{references}

\appendix

\label{lastpage}

\end{document}